\documentclass{article}
\usepackage[utf8]{inputenc}

\title{Efficient electronic cooling by niobium-based superconducting tunnel junctions}
\author{J. Hätinen*, A. Ronzani, R.P. Loreto, E. Mykkänen,\\ A. Kemppinen, K. Viisanen, T. Rantanen, J. Geisor,\\ J.S. Lehtinen, M. Ribeiro, J-P. Kaikkonen, O. Prakash,\\ V. Vesterinen, C. Förbom, E.T. Mannila, M. Kervinen,\\  J. Govenius, M. Prunnila$^{\dagger}$\\\\
*joel.hatinen@vtt.fi\\
$^{\dagger}$mika.prunnila@vtt.fi\\
VTT Technical Research Centre of Finland}

\date{}
\usepackage[a4paper, total={15cm, 20cm}]{geometry}
\usepackage{graphicx}
\usepackage{mathtools}

\begin{document}

\maketitle
\begin{abstract}
Replacing the bulky cryoliquid-based cooling stages of cryoenabled instruments by chip-scale refrigeration is envisioned to disruptively reduce the system size similar to microprocessors did for computers. Electronic refrigerators based on superconducting tunnel junctions have been anticipated to provide a solution, but reaching the necessary above the 1-K operation temperature range has remained a goal out of reach for several decades. We show efficient electronic refrigeration by Al-AlO$_x$-Nb superconducting tunnel junctions starting from bath temperatures above 2 K. The junctions can deliver electronic cooling power up to approximately mW/mm$^2$, which enables us to demonstrate tunnel-current-driven electron temperature reduction from 2.4 K to below 1.6 K (34\% relative cooling) against the phonon bath. Our work shows that the key material of integrated superconducting circuits—niobium—enables powerful cryogenic refrigerator technology. This result is a prerequisite for practical cryogenic chip-scale refrigerators and, at the same time, it introduces a new electrothermal tool for quantum heat-transport experiments.
\end{abstract}

\section*{Introduction}
The ability to control the thermodynamic temperature of matter down to fractions of a degree from the absolute zero is important for modern science and technology. Landmark scientific milestones enabled by access to a low-temperature range from the discovery of superconductivity~\cite{Onnes1911FurtherTemperatures} and quantum Hall effects \cite{Klitzing1980NewResistance, Tsui1982Two-DimensionalLimit, Chang2013ExperimentalInsulator} to the first realization of a solid-state quantum processor postulating quantum advantage~\cite{Arute2019QuantumProcessor}. Lowering of the temperature reduces thermal noise and brings in useful phases of matter, like superconductivity, and, thereby, numerous electronic and photonic devices benefit from low temperatures. Application fields utilizing cryogenic devices range from material analysis to space technology. Timely examples include different classical sensors such as x-ray calorimeters~\cite{Wilkes2022X-rayAge} and THz bolometers~\cite{Lewis2019ADetectors}, quantum information components like Josephson junction~\cite{Arute2019QuantumProcessor} and quantum-dot-based quantum processors~\cite{Philips2022UniversalSilicon}, and single-photon detectors of quantum key distribution systems~\cite{Grunenfelder2023FastSystems}. The main tool of modern cryogenics is the pulse-tube refrigerator~\cite{DeWaele2000Pulse-tubeProspects}, which allows temperatures down to 2.2~K by using $\prescript{4}{}{\mathrm{He}}$ as the cryoliquid~\cite{XuAK,Kittel2010UltimateCryocoolers}. Reaching the subkelvin regime requires an additional cooling stage that utilizes, for example, pumping of pure $\prescript{3}{}{\mathrm{He}}$ or $\prescript{3}{}{\mathrm{He}}$--$\prescript{4}{}{\mathrm{He}}$ mixture \cite{Radebaugh2009Cryocoolers:Developments}. The $\prescript{3}{}{\mathrm{He}}$--$\prescript{4}{}{\mathrm{He}}$-based dilution refrigerators provide routinely a base temperature of approximately 10~mK and they have gained vast popularity during the rise of quantum technologies~\cite{Zu2022DevelopmentReview}.

At low temperatures, the degrees of freedom of free charge carriers and lattice phonons in metals and semiconductors become so decoupled that even minute heat flows to the different subsystems can introduce large temperature differences between them~\cite{Roukes1985HotTemperatures}. This potential for overheating is utilized in many high-sensitivity detectors like in the above-mentioned cryogenic bolometers and calorimeters. Likewise, carriers can be refrigerated below the lattice temperature by removing heat by electronic energy-filtering processes. Such electronic cooling can be achieved, for example, by passing electrical current through normal metal--insulator--superconductor (\textit{N-I-S}) junctions, where thermioniclike electron cooling is obtained by matching the applied voltage to the quasiparticle energy gap in the superconducting electrode~\cite{Nahum1994ElectronicJunction, Leivo1996EfficientJunctions}. Another example is coolers based on \textit{S-I-S'} junctions, where \textit{S} and \textit{S'} are superconductors with a different energy gap \cite{Manninen1999CoolingTunneling, Quaranta2011CoolingNanorefrigerators}. Since the early proofs of concept~\cite{Nahum1994ElectronicJunction, Leivo1996EfficientJunctions}, \textit{N-I-S} components have proved to be useful for probing and manipulating thermal baths in mesoscopic and quantum heat-transport physics experiments~\cite{Giazotto2006OpportunitiesApplications}, even for resetting quantum bits~\cite{Tan2017Quantum-circuitRefrigerator}. The refrigeration of macroscopic, technologically relevant payloads~\cite{Muhonen2012Micrometre-scaleRefrigerators} has been demonstrated in the subkelvin regime by utilizing the macroscopic size to thermalize phonons into the cooled electrons and implementing phonon-segregation techniques such as membrane suspensions~\cite{Pekola1999NISRefrigeration} and lateral cold fingers~\cite{Clark2005CoolingRefrigerators}. A fully scalable approach for \textit{N-I-S} junction coolers enabling generic solid-state cooling platforms has been discovered recently in an experiment that demonstrated the refrigeration of a macroscopic silicon chip~\cite{Mykkanen2020ThermionicBlocking}. Compared to massive dilution refrigerators, such miniaturized and scalable solid-state cooler stages are envisioned to enable entirely new functionalities, such as compact three-dimensional (3D) packages of several chips at different temperatures~\cite{Hatinen2023ThermalAssemblies,Kemppinen2021CascadedLimits}.

Due to the straightforward fabrication process by evaporation and thermal oxidation that produces high-quality oxide tunnel barriers, Al has been the superconductor of choice for \textit{N-I-S} devices. However, the magnitude of its superconducting transition temperature (approximately 1~K) limits the \textit{N-I-S} device operation below approximately 500~mK. Exploitation of higher critical temperature elemental superconductors~\cite{Quaranta2011CoolingNanorefrigerators, Chaudhuri2014SuperconductingJunctions, Nguyen2016CascadeJunctions} was sought to expand the temperature range of \textit{N-I-S} elements and to eventually create scalable solid-state cooler cascades~\cite{Kemppinen2021CascadedLimits} that could bring dilution refrigerator equivalent sub-kelvin temperatures accessible without using the rare and expensive $\prescript{3}{}{\mathrm{He}}$ isotope. To this end, especially, Nb is an attractive material: it has a critical temperature of approximately 9~K and most complex integrated superconducting circuits to date utilize Nb-based \textit{S-I-S} junctions~\cite{Braginski2019SuperconductorOutlook}. However, the attempts to extend the \textit{N-I-S} device temperature range with Nb have not been successful; the utilized fabrication processes introduced too high subgap leakage and/or too low junction transparency~\cite{Capogna1997ElectronicJunctions, Buonomo2003, Nevala2012Sub-micronNb, Julin2016ApplicationsJunctions} to permit sizeable and efficient \textit{N-I-S} cooling. In contrast, in this work we report a disruptive leap in the state of the art: a modern foundry-compatible Al-AlO$_x$-Nb tunnel junction technology yielding both transparent and low-leakage barriers for use in electron refrigerators operating above 2~K. With this technology, we demonstrate significant electron refrigeration from 2.4 to below 1.6~K. Our work finally achieves the highly pursued Nb-based electrothermal element that can be utilized both in mesoscopic heat-transport science and in various applications from radiation detectors to scalable solid-state refrigerators.

\section*{On-chip refrigerator device}

\begin{figure}[t!]
    \centering
    \includegraphics[width = 12cm]{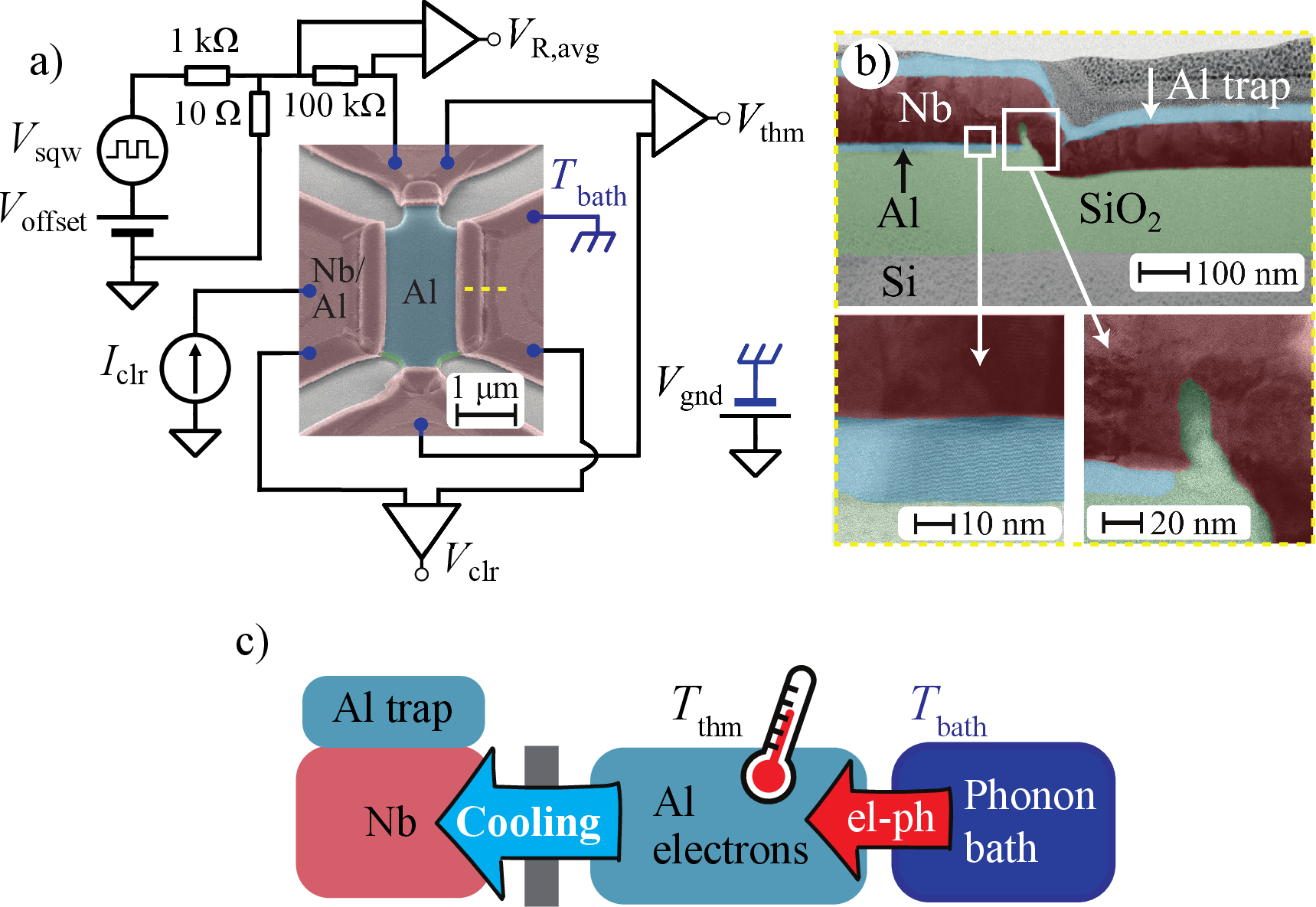}
    \caption{a) Scanning electron micrograph of the refrigerator with the experimental wiring and instrumentation. Cooled 20-nm-thick Al film is common to all four junctions. b) Cross-section transmission electron micrographs (TEM) of the junction. Tunnel barrier is located between the Al layer (light blue) on top of the SiO$_2$ (green) and Nb (dark red). Top Al layer (light blue) acts as a quasiparticle trap. SiO$_2$ spacer prevents direct galvanic contacts from the Nb to the cooled Al. Gray layers above the Al trap are protection layers deposited during the TEM lamella preparation. Cooled Al film and the spacer are shown in the magnified images on the bottom. c) Thermal model of the refrigerator. Electrons in the Al electrode are cooled by thermionic energy filtering through the tunnel oxide. Electron-phonon (el-ph) coupling in Al counteracts the cooling.}
    \label{fig:FIG1-device}
\end{figure}

Figure \ref{fig:FIG1-device} displays device micrographs, schematics of the experimental setting and heat-flow schematics. In Figs. \ref{fig:FIG1-device}a) and \ref{fig:FIG1-device}b), respectively, we show a false-colored scanning electron micrograph of the refrigerator with an experimental electronic circuit schematics and false-colored cross-section transmission electron micrographs of the tunnel junction. The cross-section location is indicated by the dashed yellow line in Fig. \ref{fig:FIG1-device}a). The device has a 20-nm-thick Al film on oxidized Si substrate common to four junctions connected to it. Two larger junctions (2.8~µm$^2$ each, referred as cooler later in the text) on the left and the right of the refrigerator are used for cooling; the smaller junction on the top (0.49~µm$^2$) is used as a thermometer and the symmetric junction on the bottom provides an independent probe for the electrical potential of the Al electrode.

The refrigerator was fabricated on a 150-mm \textit{p}-type Si wafer with 200-nm-thick thermally grown SiO$_2$, using a variant of side-wall spacer passivated structure (SWAPS) fabrication process, which was originally developed for Nb-Nb Josephson junctions~\cite{Gronberg2017Side-wallProcess}, and has been used, e.g., for the scalable fabrication of Josephson parametric amplifiers~\cite{PerelshteinBroadbandMetamaterial} and qubits~\cite{Anferov2024ImprovedQubits}. A trilayer with Al~(20~nm)-AlO$_x$-Nb~(60~nm) was deposited \textit{in-situ} with dc sputtering and patterned with ultraviolet stepper lithography and reactive ion etching. A side-wall spacer passivation layer of SiO$_2$ was deposited by plasma-enhanced chemical vapor deposition (PECVD) and plasma etched to form the SWAPS ~\cite{Gronberg2017Side-wallProcess} to prevent direct galvanic contacts from the bottom Al electrode on the trilayer to the wiring layer of Nb (100 nm)-Al (30 nm) deposited on top. SWAPS spacer and device cross section are displayed in the panels of Fig. \ref{fig:FIG1-device}c). Prior to depositing the wiring layer, argon milling was performed to ensure a galvanic contact between the trilayer top Nb and the Nb wiring layer. The 30-nm-thick Al top-layer was preventively designed to act as a quasiparticle trap to enhance the cooler performance \cite{Agulo2004EffectiveTraps, Nguyen2013TrappingCooler}. 

A key figure of merit in the quality of superconducting tunnel-junction processes is the normalized subgap leakage, typically parametrized through the phenomenological Dynes parameter \cite{Dynes1978DirectSuperconductor}. A set of simple cross-wire tunnel junctions was fabricated with a dedicated diagnostic stack (see Sec. S1 within the Supplemental material~\cite{supplementary}). Based on measurements of their differential conductance, the typical superconducting energy gap in Nb was found to be approximately equal to 1.44~meV in the operative range of interest, with relative sub-gap leakage lower than 10$^{-3}$, ideally suited to electron-cooling applications.

Figure~\ref{fig:FIG1-device}c) shows the thermal model for the refrigerator. The electron gas in the Al electrode is cooled by thermionic energy filtering of the tunneling electrons through AlO$_x$ to Nb \cite{Giazotto2006OpportunitiesApplications}. The electron system is coupled to the surrounding phonon bath, resulting in $T^5$-dependent electron-phonon heat flow when Al is in the normal state~\cite{Giazotto2006OpportunitiesApplications}. The Al phonon bath is further connected to the silicon chip and the sample stage at the cold head of the cryostat, but due to the large contact areas, all aforementioned phonon baths are assumed to be in equilibrium. Thus the relevant heat flows in the refrigerator are from the electron cooling and the counteracting electron-phonon coupling. The temperature of the Al electron system $T_\textrm{thm}$ was determined by junction thermometry (see below).

\section*{Electrothermal performance}

%% IV measurement scheme first, then IV data, then cooling measurement, then cooling data
The refrigerator was characterized in a cryostat first by dc biasing the cooler junctions and recording the corresponding voltage to find the regions of interest in the current-voltage (\textit{I-V}) characteristics. The junctions were biased with current sources due to the low resistance of the junctions (from below 1~$\Omega$ to 1.5~k$\Omega$ as a function of temperature) compared to the line resistances in the cryostat (approximately 0.3~k$\Omega$ per line). Figure \ref{fig:FIG1-device}a) shows the cooler dc bias as $I_\textrm{clr}$ and the voltage $V_\textrm{clr}$ was measured using a preamplifier in differential mode. The amplified voltage was recorded using a digital multimeter. The sample was grounded through the junction on the right side of the sample to the sample holder mounted on the low-temperature part of the cryostat.

\begin{figure}[t!]
    \centering
    \includegraphics[width = 14cm]{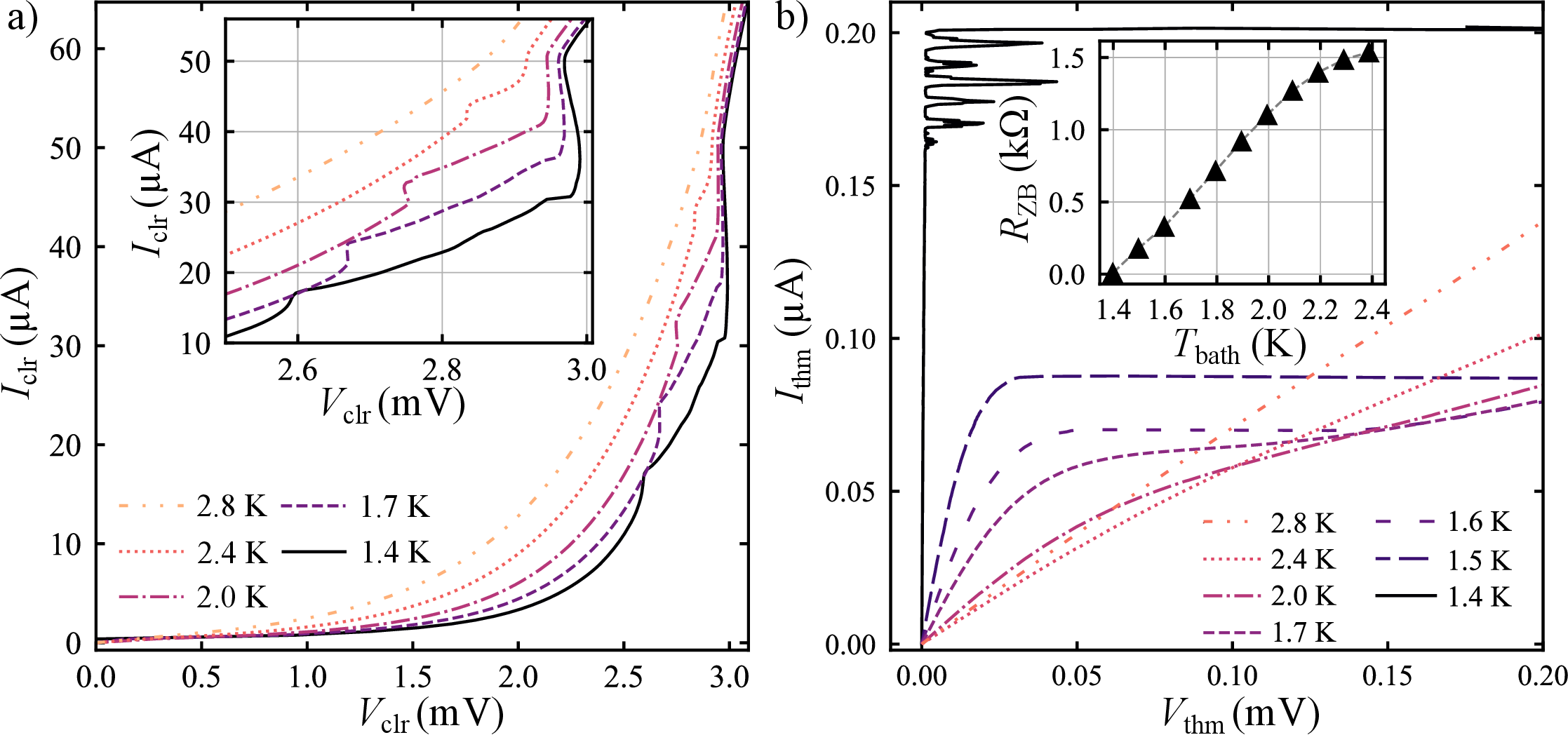}
    \caption{a) Cooler \textit{I-V} characteristics at several $T_\textrm{bath}$. \textit{N-I-S}-like characteristics are seen at $T_\textrm{bath}$~=~2.8~K and above, while features with negative differential resistance appear at lower bath temperatures, emphasized in the inset. b) Thermometer \textit{I-V} characteristics near zero bias at the same $T_\textrm{bath}$ as in a), with an addition of $T_\textrm{bath} =$ 1.5 and 1.6 K. Inset: zero-bias resistance of the thermometer $R_\textrm{ZB}$ as a function of $T_\textrm{bath}$. The dashed line is an interpolant between the data points, used for converting recorded $R_\textrm{ZB}$ to equivalent electronic temperature under active cooler operation.}
    \label{fig:FIG2-IVdata}
\end{figure}

%%Cooler IV data
The measured cooler \textit{I-V} characteristics at several bath temperatures $T_\textrm{bath}$ are shown in Fig.~\ref{fig:FIG2-IVdata}a) up to $V_\textrm{clr}$~=~3~mV. The specific resistance of the junctions was measured to be around 17~$\Omega~$µm$^2$. At the bath temperature $T_\textrm{bath}$~=~2.8~K, the cooler \textit{I-V} characteristics resemble that of a \textit{N-I-S} junction i.e. the Al electrode is in normal state and Nb is superconducting \cite{Giazotto2006OpportunitiesApplications}. Instead, at lower $T_\textrm{bath}$ the \textit{I-V}s start to develop two features characterized by negative differential resistance (NDR) between cooler voltages $V_\textrm{clr}$~=~2.5 and 3~mV. The feature present at $V_\textrm{clr} \approx $ 3~mV is observed to shift to lower values on voltage axis with increasing $T_\textrm{bath}$. In contrast, the feature present at lower voltages ($V_\textrm{clr}$~=~2.5~mV~to~2.85~mV) is shifting to lower values with decreasing temperature. We note that the cooler is a series of two Nb-AlO$_x$-Al junctions, where the thin Al is superconducting below 1.38~K \cite{Chubov1969DependenceFilms} (see Sec. S2 within the Supplemental Material~\cite{supplementary}).

%%Thermometer measurement scheme
The thermometer \textit{I-V}s were recorded by utilizing a method suited to eliminate the voltage drift from the preamplifier to accurately resolve the µV-level voltages for the \textit{I-V}s near zero bias (ZB). High resolution and voltage drift-free measurement allowed us to measure temperature close to zero-bias regime, thus neglecting self-heating effects from the thermometer. The measurement scheme is shown in Fig.~\ref{fig:FIG1-device}a) by lines connected to the smaller junctions at the top and the bottom of the refrigerator. The thermometer was biased with a square-wave excitation with a frequency of 200~Hz and amplitude varying between zero and $V_\textrm{sqw}$. A voltage divider with a ratio of 100 (1 k$\Omega$ in series and 10 $\Omega$ to the ground) was used to reduce the supply voltage to the desired magnitude and a bias resistor of 100~k$\Omega$ was used to define current bias to the thermometer. Current was acquired by measuring the voltage drop $V_\textrm{R, avg}$ over the bias resistor with a digital multimeter after amplifying the signal with a voltage preamplifier in differential mode. Current was then calculated with $I_\textrm{thm}$~=~$2 \times V_\textrm{R, avg}/102.6$, where 2 is used to get the peak value of the signal from the average value output from the digital multi-meter and 102.6 is the calibrated gain of our voltage preamplifier (NF Corp. model LI-75A). Due to a small voltage offset ($V_\textrm{gnd}\sim$ 5~mV) originating from thermoelectric effects in the cryostat, a dc voltage component $V_\textrm{offset}$ was used to compensate for it (its value set by requiring zero current sensed across the 100-k$\Omega$ resistor when $V_\textrm{sqw}=0$). The thermometer voltage $V_\textrm{thm}$ was measured in parallel by first amplifying the signal with a preamplifier in differential mode and then reading the signal with a lock-in amplifier. An oscilloscope monitoring the lock-in input channel was used to confirm the bias and response waveforms and magnitudes to match the expected values.

%%Thermometer IV data
The thermometer \textit{I-V}s near ZB with the cooler inactive \textit{i.e.} cooler bias $I_\textrm{clr}~=~0$ were recorded at several $T_\textrm{bath}$ are shown in Fig.~\ref{fig:FIG2-IVdata}b). At $T_\textrm{bath}$~=~1.4~K, the \textit{I-V} shows supercurrent-like behavior until a threshold current of 0.20~µA is exceeded and the voltage abruptly increases, switching the thermometer to resistive state. Importantly, at temperatures above 1.4~K, we do not observe high subgap tunnel resistance values that decline with increasing temperature, which would be expected from conventional \textit{N-I-S} tunnel transport theory \cite{Giazotto2006OpportunitiesApplications}. In contrast, all realized junctions display anomalously high ZB conductance until a temperature-dependent threshold bias current is reached. This phenomenology appears analogous to the one shown in transparent \textit{N-I-S} junctions \cite{Hekking1993InterferenceSuperconductor, Hekking1994SubgapInterface, Faivre2015AndreevThermometry}. Overall, the ZB resistance increases linearly with temperature in range 1.4~K to 2.0~K, and monotonically until saturation at 2.4~K (see inset of Fig.~\ref{fig:FIG2-IVdata}b) ). In the following, we use this mechanism for thermometry. The ZB resistances $R_\textrm{ZB}$ were extracted with a linear fit to the data points below 9-nA bias currents $I_\textrm{thm}$. In this current range, the bias applied to the thermometer corresponds to a maximum power of 1.3~pW at $T_\textrm{bath}$~=~2.4~K. $R_\textrm{ZB}$ at each $T_\textrm{bath}$ are shown in the inset of Fig.~\ref{fig:FIG2-IVdata}b) from 1.4~K to 2.4~K with a linear interpolation line between the data points used as the thermometer calibration. The operative limitations of this thermometer are discussed in detail in the Sec. S3 within the Supplemental Material~\cite{supplementary}.

%% Cooling measurement
To characterize the cooling performance of the refrigerator, the two measurement schemes discussed above were combined, thus utilizing the full circuit shown in Fig. \ref{fig:FIG1-device}a). The thermometer \textit{I-V}s near ZB were recorded as a function of $I_\textrm{clr}$ at several $T_\textrm{bath}$. The cooler bias points were chosen based on the \textit{I-V}s shown in Fig. \ref{fig:FIG2-IVdata}a).

\begin{figure}[h!]
    \centering
    \includegraphics[width = 14 cm]{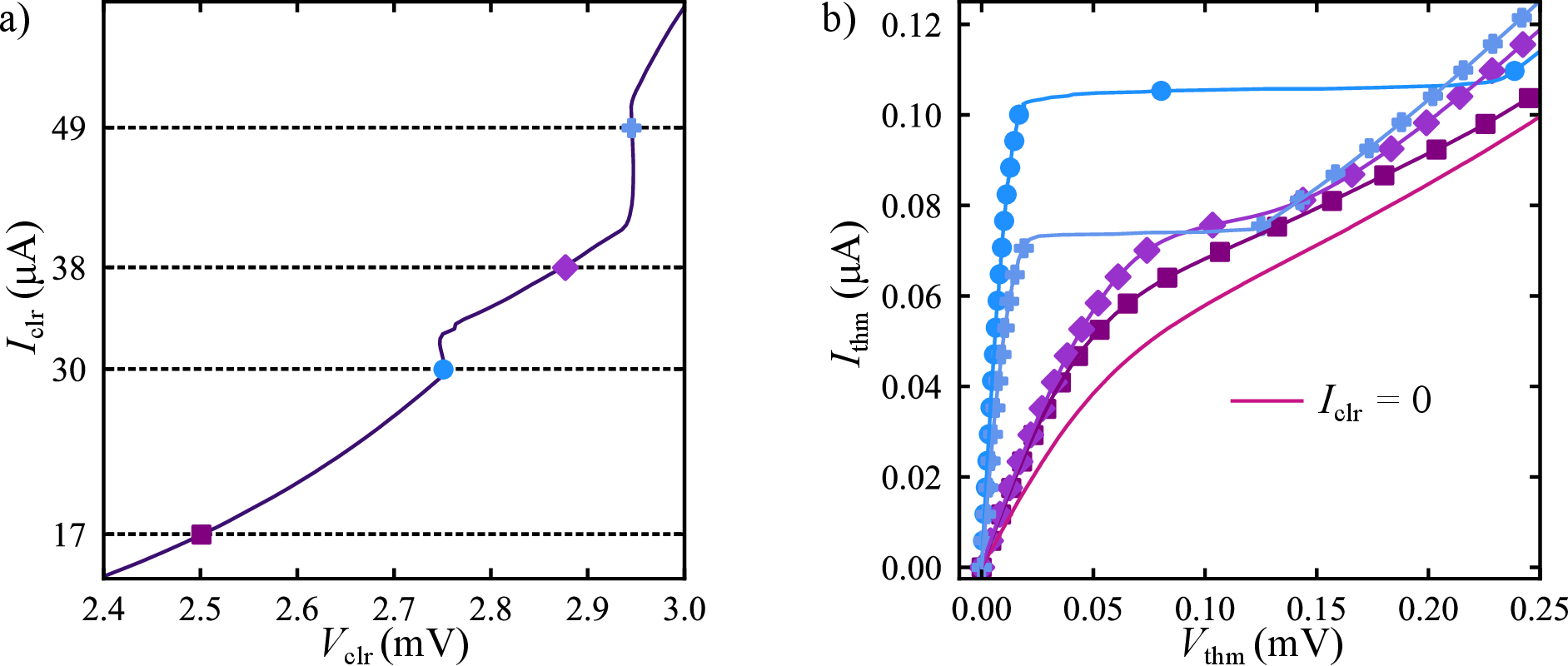}
    \caption{a) Cooler \textit{I-V} between $V_\textrm{clr}$~=~2.4 and 3.0~mV with different current bias points marked at 2.0~K. b) Thermometer \textit{I-V}s near zero bias at 2.0~K when biasing the cooler at the points indicated by the markers in panel a. Solid lines represent the data and the markers serve as a visual guide. Zero-bias resistance of the thermometer is decreased by biasing the cooler compared to the zero cooler bias. Additionally, the threshold currents are significantly enhanced by biasing the cooler at the optimal cooling bias points. Both observations indicate cooling.}
    \label{fig:FIG3-thermoData}
\end{figure}

%% Cooling data
Biasing the cooler at the points of interest in the \textit{I-V} shown at Fig.~\ref{fig:FIG3-thermoData}a) alters the thermometer response compared to the case when the cooler is inactive, \textit{i.e.} $I_\textrm{clr}$~=~0, as shown in Fig.~\ref{fig:FIG3-thermoData}b). At $T_\textrm{bath}$~=~2.0~K, biasing the cooler at the given points decreases $R_\textrm{ZB}$ significantly when operating the cooler at the NDR features at $V_\textrm{clr}$~=~2.5 and 2.93~mV. We also observe higher threshold current, similarly to that seen in $T_\textrm{bath}$~=~1.5 and 1.6~K with the cooler inactive (Fig.~\ref{fig:FIG2-IVdata}b). Both decreased $R_\textrm{ZB}$ and increased threshold current suggest that the electron temperature in the Al film has decreased by the operation of the cooler junctions. Similar analysis was done at $T_\textrm{bath}$~=~1.4 to 2.4~K (see Sec. S4 within the Supplemental Material~\cite{supplementary}).

\begin{figure}[h!]
    \centering
    \includegraphics[width = 14cm]{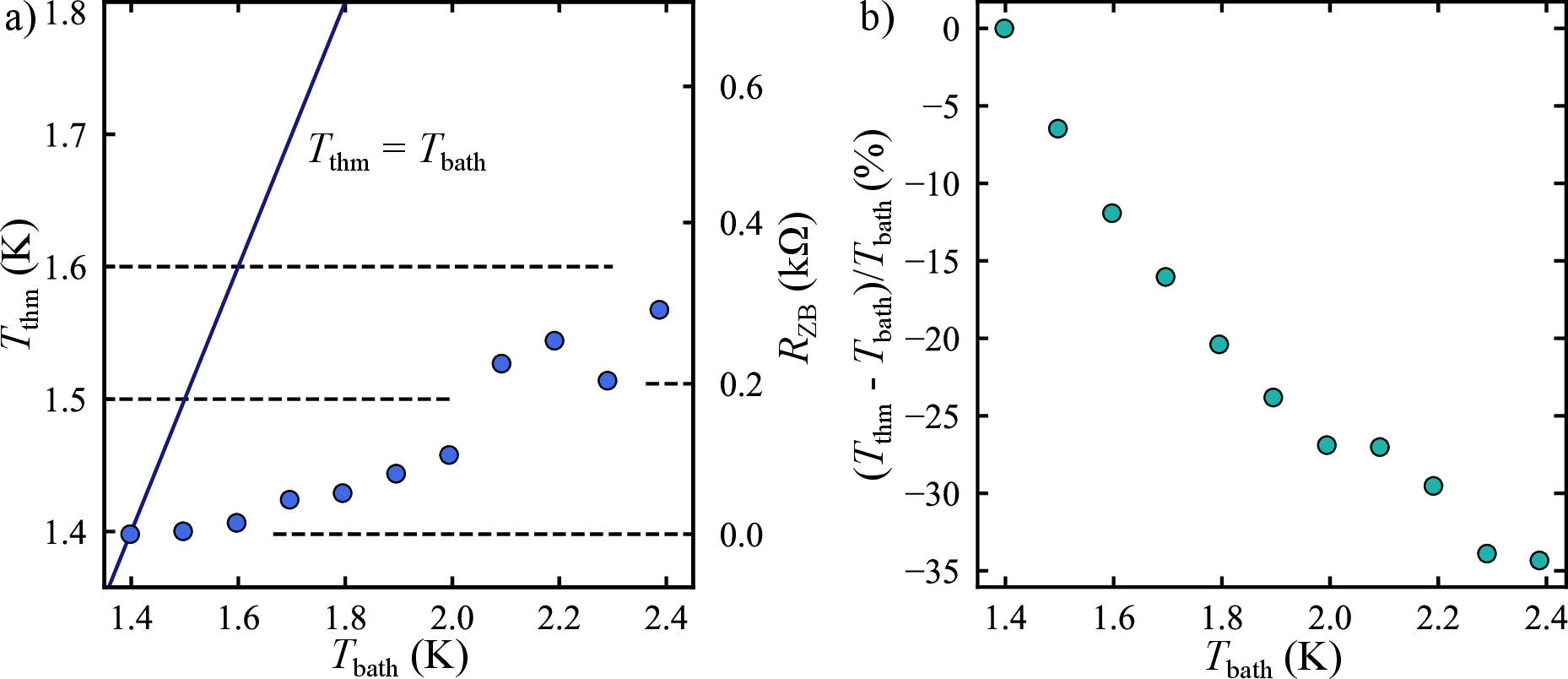}
    \caption{a) Extracted $R_\textrm{ZB}$ (right \textit{y} axis) of the thermometer and the conversion to temperature $T_\textrm{thm}$ (left \textit{y} axis) as a function of $T_\textrm{bath}$. We observe cooling from all bath temperatures to below 1.6~K. Calibration of the thermometer has a lower limit of 1.4 K, thus the $R_\textrm{ZB}$ values at $T_\textrm{bath}$~=~1.4 and 1.5~K convert to $T_\textrm{thm}$~$<$~1.4~K. The solid line is a visual guide for $T_\textrm{thm} = T_\textrm{bath}$. b) Relative temperature difference ($T_\textrm{thm}$ - $T_\textrm{bath}$)/$T_\textrm{bath}$ as a function of $T_\textrm{bath}$. Maximum cooling of 34\% is observed from $T_\textrm{bath}$~=~2.4~K.}
    \label{fig:fig4-mainresult}
\end{figure}

%% Discussion and main results
Figure \ref{fig:fig4-mainresult} captures the main result of the paper. The extracted $R_\textrm{ZB}$ values of the thermometer at the best found cooler bias values are shown as a function of $T_\textrm{bath}$ on the right-side \textit{y} axis of Fig. \ref{fig:fig4-mainresult}a). The thermometer calibration was used to convert the $R_\textrm{ZB}$ values to the temperature $T_\textrm{thm}$ which is shown on the left-side \textit{y} axis of Fig. \ref{fig:fig4-mainresult}a). We see cooling to below 1.6~K from all investigated $T_\textrm{bath}$ values. Maximum absolute cooling is observed from $T_\textrm{bath}$ = 2.4 to 1.57~K corresponding to a relative cooling of 34\% from the phonon bath (see Fig. \ref{fig:fig4-mainresult}b)). Importantly, through optimal cooler junction bias we can also observe the appearance of a sizeable ($>$~100~nA) superconducting branch at bath temperature values as high as 1.6~K, as well as enhance by more than 400\% the supercurrent branch present at $T_\textrm{bath}$~=~1.4~K. These effects, shown in Fig.~S5 within the Supplemental Material \cite{supplementary}, are consistent with and corroborate the above interpretation of cooling based on the ZB resistance thermometry.

The NDR feature and its temperature dependence observed near the energy gap edge at $V_\textrm{clr} \approx$ 3~mV can be attributed to overheating of the Nb electrode due to the large power injected to the superconductor at high biases \cite{Nguyen2016CascadeJunctions}. Instead, the NDR feature at smaller $V_\textrm{clr}=~2.5~\mathrm{to}~2.85~\textrm{mV}$ can be qualitatively interpreted as the emergence of superconductivity in the Al volume beneath the junction. This manifests with the appearance of a local bias-dependent gap $\Delta_\textrm{Al}$, that leads to divergent differential conductance at junction voltages equal to $2(\Delta_\textrm{Nb} - \Delta_\textrm{Al})/e$, where $\Delta_\textrm{Nb}$ is the generally bias-dependent energy gap of Nb and $e$ is the elementary charge. This phenomenon is commonly observed in \textit{S}$_1$\textit{-I-S}$_2$ systems at finite temperature \cite{Ronzani2014HighlySensitive}. Notably, the smaller the temperature of Al, the larger $\Delta_\textrm{Al}$ becomes, leading to the observed shift to smaller $V_\textrm{clr}$ with decreasing bath temperature, mentioned above. Crucially, the interpretation of Al volume becoming superconducting is supported by the simultaneous observation of the supercurrent through the thermometer at $T_\textrm{bath} > 1.4$~K (Fig.~S5 within the Supplemental Material \cite{supplementary}).

The refrigerator was simulated with a finite-element method to estimate the temperature at the cooler junctions and the corresponding cooling power density (see Sec. S5 within the Supplemental Material~\cite{supplementary}). We assumed normal-state Al in calculating the thermal conductivity and electron-phonon coupling \cite{Santhanam1984InelasticFilms}. The simulations show that the temperature across the 20-nm-thick Al film is uniform up to 1.5\%. The estimated cooling power density is in the order of $p\sim$~mW/mm$^2$. In comparison, in Ref.~\cite{Quaranta2011CoolingNanorefrigerators} the authors report $p=65$~µW/mm$^2$ in a V-based refrigerator operating at 1~K and in Ref.~\cite{Nguyen2013TrappingCooler} the reported estimate for an Al-based refrigerator operating at 0.3~K is $p=0.7$~µW/mm$^2$. Remarkably, we reach cooling power similar to that of a thermionic refrigerator based on asymmetric double-barrier semiconductor heterostructures operating at room temperature \cite{Yangui2019CoolingHetero}. 

Cooling power of superconducting tunnel-junction refrigerator scales both as a function of temperature and the superconductor energy gap. In the optimal case, the gap is matched for the operating temperature ($\Delta\propto T$), which results in cooling power density $\propto T^2$~\cite{Mykkanen2020ThermionicBlocking,Kemppinen2021CascadedLimits}. An approximative comparison for the technical performance of different of cooler junctions operating at different temperatures can thus be obtained by comparing the values of $p/T^2$ for each case, which yields about 8~µW/mm$^2$K$^2$, 65~µW/mm$^2$K$^2$, and 280~µW/mm$^2$K$^2$ for the above-mentioned Al- and V-based devices, and our refrigerator operating at 2.4~K, respectively. This indicates that our cooler is efficient compared to other electronic refrigerators even if the larger temperature and superconducting gap are taken into account. This is also an indication of high-quality junctions in terms of transparency and leakage current, achieved with our foundry-compatible Nb-based fabrication process.

\section*{Conclusions}

In conclusion, we have demonstrated significant cooling of free electrons with respect to lattice phonons up to 2.4-K bath temperature. At 2.4~K, we observe cooling of electrons down to 1.6~K – equivalent to 34\% relative cooling. The temperature range was limited by the electron thermometry, \textit{i.e.}, the cooling effect is expected to work similarly also above 2.4~K. The cooling result was enabled by developing Nb-based superconducting tunnel-junction refrigerator technology. With our modern foundry-compatible Al-AlO$_x$-Nb tunnel junctions, we obtained a cooling power density of approximately mW/mm$^2$, which is expected to enable substantial cooling in practical chip scale refrigerators. Most significantly, after decades of research on tunnel-junction-based electronic coolers, here we demonstrated a device that can operate at the base temperature of $\prescript{4}{}{\mathrm{He}}$ pulse tube refrigerators, thus paving the way for replacing bulky dilution refrigerators by miniaturized electronic coolers that operate without rare and expensive $\prescript{3}{}{\mathrm{He}}$.

\section*{Acknowledgements}
The authors thank Sari Ahlfors, Manika Maharjan and Kaisa Välimaa for technical help in the sample fabrication in VTT and OtaNano Micronova cleanroom facilities. The authors also thank Francesco Giazotto and Elia Strambini for useful discussions. 

The research was funded by the European Union’s Horizon RIA, EIC and ECSEL programmes under grant agreements No. 766853 EFINED, No. 824109 European Microkelvin Platform (EMP), No. 101113086 SoCool, No. 101007322 MatQu, and No. 101113983 Qu-Pilot. We also acknowledge financial support of Research Council of Finland through projects No. 322580 ETHEC, No. 356542 SUPSI, the QTF Centre of Excellence project No. 336817, and Business Finland through Quantum Technologies Industrial (QuTI) No. 128291 and Technology Industries of Finland Centennial Foundation and Chips JU project Arctic No. 101139908.

\section*{Conflict of interest}
The authors have no conflicts of interest to disclose.

\section*{Author contributions}
Research was conceptualized by J.H, A.R, A.K, J.S.L and M.P. Device design was done by J.H, A.R, R.P.L and K.V. Device fabrication was led by R.P.L with the support from J.H. T.R did device validation through SEM imaging. Device characterization was done in room temperature by J.H and J.Ge, and in low temperature by J.H and A.R. Data was analyzed by J.H, A.R, E.M and A.K. Finite-element-method simulation was done by E.M and A.R. with the support from J.H. Al-AlO$_x$-Nb junction development was done by R.P.L, M.R, J-P.K, O.P, V.V, C.F, E.T.M, M.K, M.P and J.G, who also supervised the junction development. M.P supervised the refrigerator work. J.H and M.P wrote the manuscript with contributions from all other authors.

\section*{Data availability}
The data that support the findings of this study are available from the corresponding author upon reasonable request.

\bibliographystyle{abbrv}

\end{document}

\maketitle

\clearpage
\section{Quality of superconducting gap in Nb electrodes}
\label{Supplementarysection: DOS}

The NIS cooling efficiency, based on energy-filtering principles, is strongly dependent on the quality of the energy gap in the superconducting electrode. To assess the suitability of our Al/Nb junction process, we fabricated and characterized diagnostic cross-wire junctions with a material stack 
comparable with the one adopted for the cooler devices considered in the main text. The differences are: 1) moderately higher junction opacity ($\approx 55~\Omega ~ $µ$\mathrm{m}^2$), 2) significantly thicker base Al layer ($\approx 100~$nm) and 3) essentially semi-infinite lateral volume of the Al part.
Compared to the situation for the cooler devices, these choices limit potential temperature differences between junction's Al quasiparticles and phonons; in the following we assume exact and uniform quasiparticle thermalization to the cryostat temperature $T=T_\mathrm{bath}$ for both electrodes.

\begin{figure}[h!]
    \centering
    \includegraphics[width = 11cm]{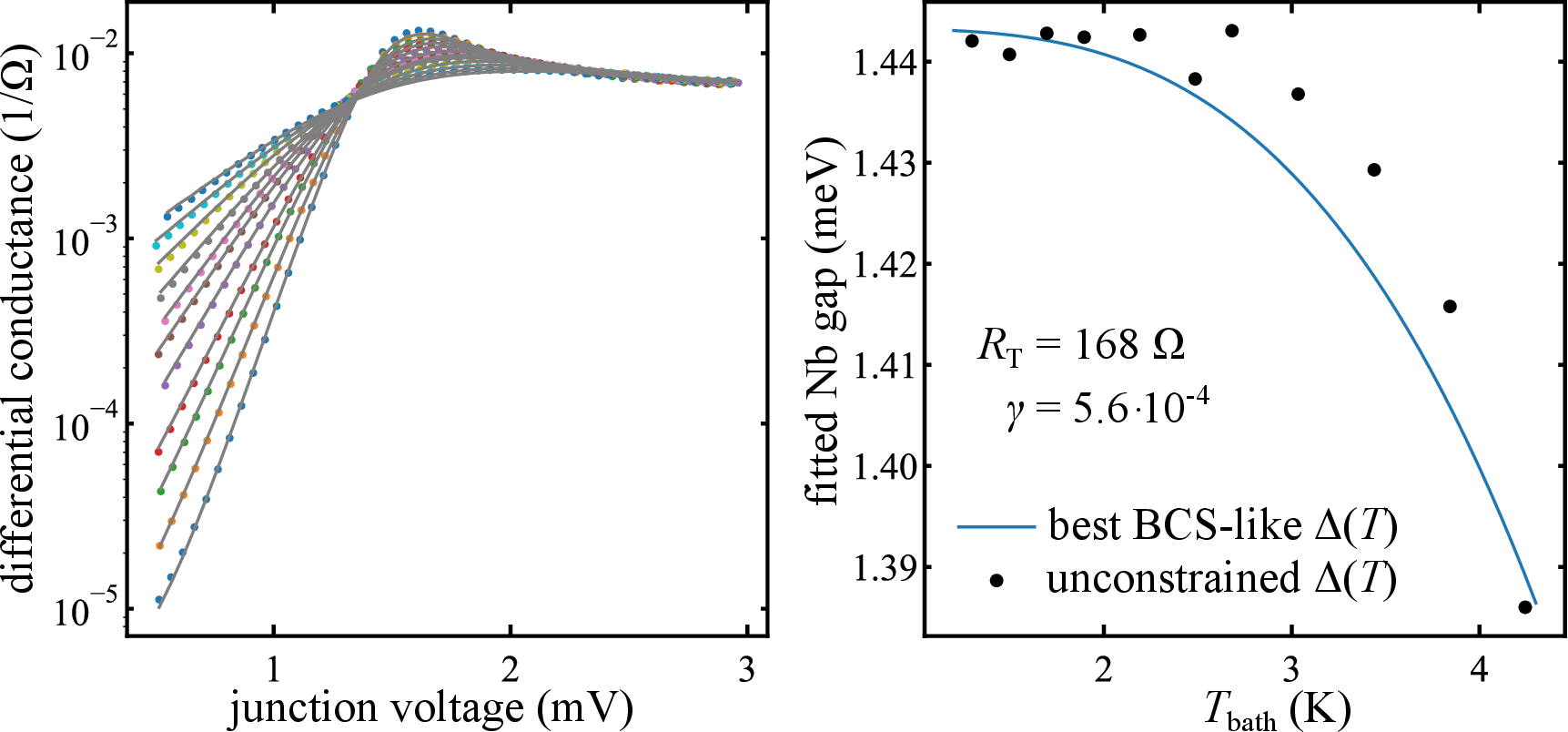}

    \caption{Assessment of effective quasiparticle
    density of states for a square cross-wire Al/AlO$_\mathrm{x}$/Nb junction of nominal width of 800 nm (typical real width $\approx 580$~nm). (a) Experimental differential conductance data (dots) at various bath temperatures (1.3, 1.5, 1.7, 1.9, 2.2, 2.5, 2.7, 3.0, 3.4, 3.8, 4.2 K) and the fitted numerical model (lines) based on conventional tunnel transport integrals.
    (b) Summary of fit parameter estimates, yielding tunnel resistance $R_T \approx 168~\Omega$, Dynes parameter $\gamma < 10^{-3}$ as well as the individual unconstrained superconducting gap amplitude values $\Delta$ (black circles) for each bath temperature. For reference, a blue line shows
    the best-fit $\Delta(T_\mathrm{Bath})$ under more stringent constraint of BCS-like gap dependence.}
    \label{fig:supplementary_DOS}
\end{figure}

Due to the presence of zero-bias anomaly, eventually developing
a \textit{S-I-S} supercurrent branch at low enough temperature, we fit experimental $dI / dV$ data in the voltage range $V \in \left[ 0.5,~3.0 \right]$~mV (Figure~\ref{fig:supplementary_DOS}). With Al in normal state, the numerical model is obtained by differentiating the current $
I_\mathrm{qp}(V, T)$ w.r.t. voltage:
$$
I_\mathrm{qp}(V, T) = \frac{1}{e~R_T} \int_{-\infty}^{+\infty} \mathrm{d}E~\rho (E+eV) \left[ \frac{1}{1+\exp [E/(k_B T)]} - \frac{1}{1+\exp [(E+eV)/(k_B T)]} \right] \quad ,
$$
where the Nb gap is parametrized by the Dynes density of states, defined as:
$$
\rho(x) = \left| \mathcal{R} \left[ \frac{x/\Delta + \mathrm{i}\gamma}{\sqrt{(x/\Delta + \mathrm{i}\gamma)^2-1}} \right] \right| \quad .
$$
In the latter, low values of $\gamma$ ($\approx 10^{-3}$ or better) are associated with high-quality superconducting gap and result in efficient \textit{N-I-S} cooling, \textit{i.e.,} unimpeded by sub-gap leakage.

The numerical fit is performed by least-square minimization of residuals with uniform weights in logarithmic scale. $R_\mathrm{T}$ and $\gamma$ are global parameters for the full dataset (figure~\ref{fig:supplementary_DOS}(a), supplemented by
the effective $\Delta$ for each bath temperature. The results are summarized in figure~\ref{fig:supplementary_DOS}b, where we can appreciate not only energy gap amplitude persisting at 1.44~meV until $T_\mathrm{bath}=3$~K, but especially $\gamma < 10^{-3}$, a value remarkably low considering the high transparency of these
junctions and ideally suited to observe sizeable \textit{N-I-S} cooling.

\section{$R(T)$ of 20-nm-thick Al film}
\label{supplementarysection:Alfilm}
\begin{figure}[h!]
    \centering
    \includegraphics[width = 8.5cm]{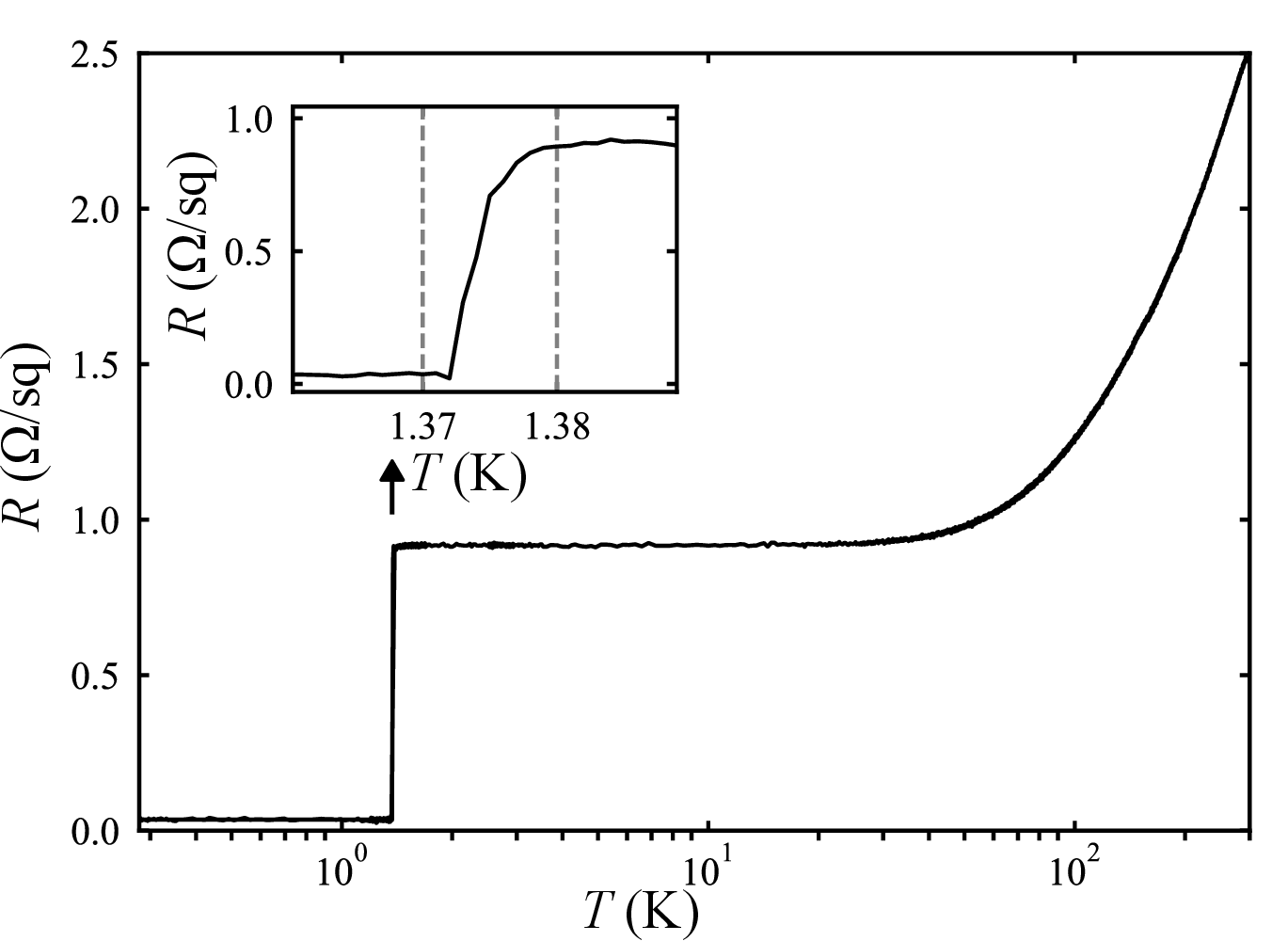}
    \caption{Sheet resistance of the 20-nm-thick Al film measured in a temperature range of 0.3 to 300~K. Superconducting transition is observed at 1.38~K. The sheet resistance is 0.9~$\Omega$/sq just above the transition temperature.}
    \label{fig:supplementary_Altransition}
\end{figure}

Figure \ref{fig:supplementary_Altransition} shows the sheet resistance of the 20-nm-thick Al film measured with a lock-in amplifier (sine wave excitation with a frequency of 20~Hz and an amplitude peak of 1~µA). The sheet resistance is around 0.9~$\Omega$/sq just above the superconducting transition at 1.38~K. The measurement was performed to have more understanding on the temperature of the Al when we observe supercurrent in the junctions sharing the common 20-nm-thick Al part. We note that due to our junctions being highly transparent (17~$\Omega~$µ$ $m$^2$), the Nb layer on top the junction might increase the critical temperature of the Al at the junction by superconducting proximity effect. The critical temperature of Al is also highly sensitive to the thickness of the film \cite{Chubov1969DependenceFilms}.

\clearpage
\section{Thermometer operation regimes}
\label{Supplementarysection: thermometer}
\begin{figure}[h!]
    \centering
    \includegraphics[width = 5 cm]{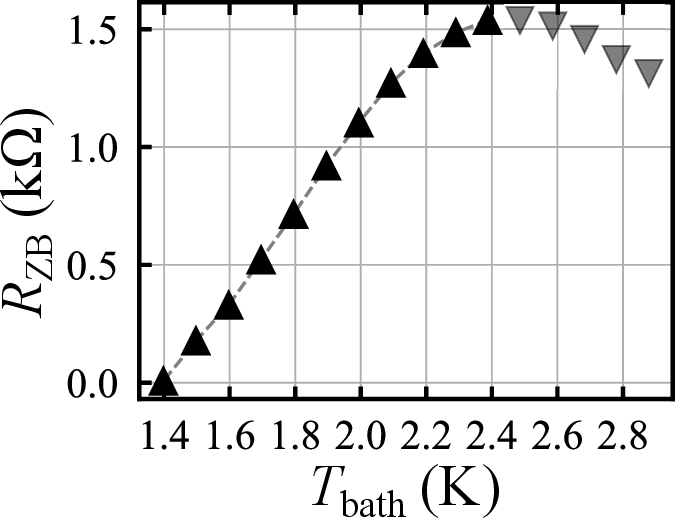}
    \caption{Thermometer zero bias resistance $R_\mathrm{ZB}$ at $T_\mathrm{bath}$~=~1.4 to 2.8~K measured with the method described in the supplementary section \ref{Supplementarysection: thermometer}. $R_\mathrm{ZB}$ reaches the maximum value between $T_\mathrm{bath}$~=~2.4 and 2.5~K. The dashed line up to 2.4~K is the linear interpolation used for thermometer calibration.}
    \label{fig:supplementary_thermometerfullcalibration}
\end{figure}

The thermometer used in the work was calibrated to a temperature range of 1.4 to 2.4~K where the zero bias resistance $R_\mathrm{ZB}$ has a monotonous behavior as a function of the bath temperature $T_\mathrm{bath}$. Here we present additional data of the thermometer at temperatures above 2.4~K, and at some intermediate to high cooler bias values. Figure \ref{fig:supplementary_thermometerfullcalibration} shows $R_\mathrm{ZB}$ as a function of $T_\mathrm{bath}$ up to 2.8~K. It is observed that $R_\mathrm{ZB}$ saturates between 2.4 and 2.5~K and decreases at increasing temperatures. In principle, there exists two possible values for temperature at a single $R_\mathrm{ZB}$ value. However, as seen in the main text Fig. 2(b), the thermometer \textit{I-V} near zero bias has a distinctive feature at $T_\mathrm{bath}$~=~1.4, 1.5 and 1.6~K, where the \textit{I-V} shows an abrupt jump in voltage after exceeding a temperature dependent threshold value of current. Similar features are not observed at any $T_\mathrm{bath}$~$>$~2.4~K, thus we conclude our thermometer provides a reliable mapping of $R_\mathrm{ZB}$ on a temperature range of 1.4 to 2.4~K.

\begin{figure}[h!]
    \centering
    \includegraphics[width = \textwidth]{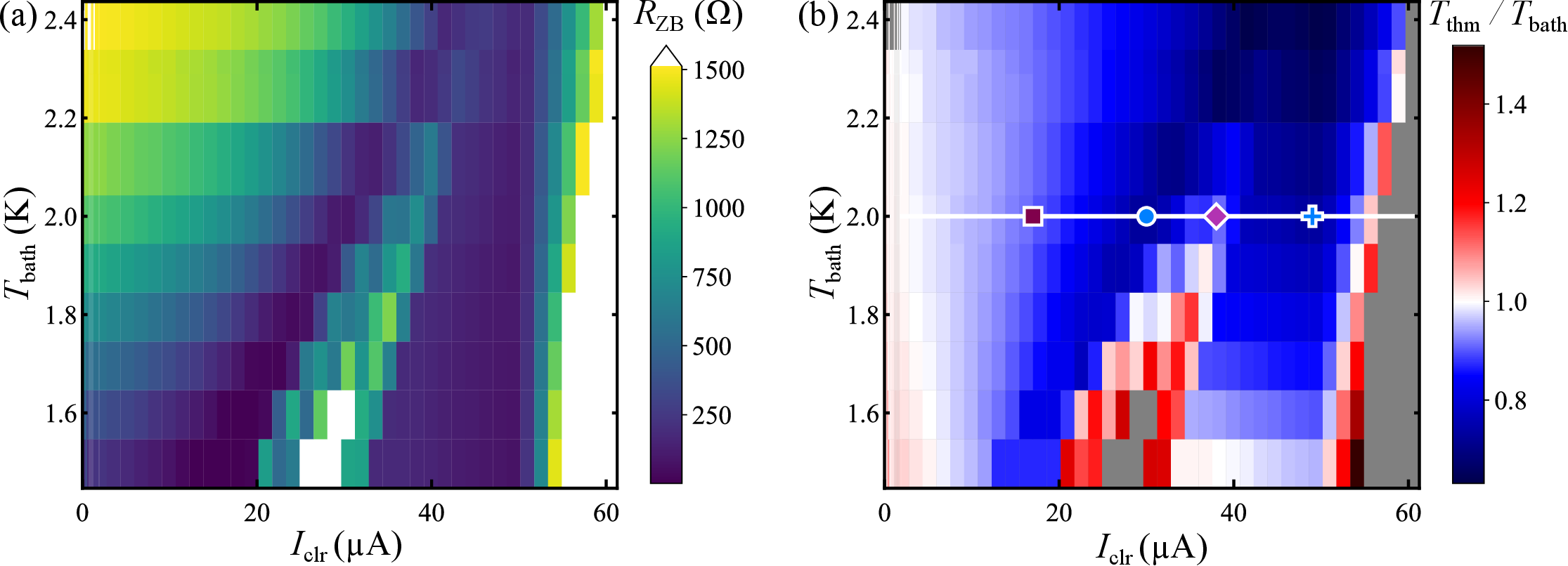}
    \caption{(a) Zero-bias resistance $R_\mathrm{ZB}$ of the thermometer and (b) relative temperature $T_\mathrm{thm}/T_\mathrm{bath}$ as a function of cooler bias $I_\mathrm{clr}$ and bath temperature $T_\mathrm{bath}$. The points above 2.1~K in (a) are linearly interpolated at each $I_\mathrm{clr}$ due to different measurement grid size below and above 2.1~K. White areas in (a) indicate higher $R_\mathrm{ZB}$ values than observed in the thermometer calibration and thus temperature in grey areas in (b) are undefined. Marked points in (b) correspond to Fig. 3(a) in the main text.}
    \label{fig:supplementary_heatmaps}
\end{figure}

Additionally, we observe $R_\mathrm{ZB}$ values of the thermometer exceeding any of the values obtained at the calibration curve at some cooler operating points. This occurs when biasing the cooler at points between the two negative differential resistance features in the cooler \textit{I-V} at the lowest $T_\mathrm{bath}$, or at biases over $I_\mathrm{clr}\sim$~55~µA, as shown in Fig. \ref{fig:supplementary_heatmaps}(a) in white. In Fig.~\ref{fig:supplementary_heatmaps}(b), we show the temperature obtained from the calibration divided by $T_\mathrm{bath}$, with the bias points used in Fig.~3(a) of the main text indicated with markers. Out of range $R_\mathrm{ZB}$ values result in undefined temperature in the grey areas in Fig.~\ref{fig:supplementary_fulldata1.4to1.6}(a) when operating the cooler at $T_\mathrm{bath}$~=~1.5~K and $I_\mathrm{clr}$ = 26, 30 and 55~µA. At these cooler bias points, the thermometer \textit{I-V}s show shapes not observed at any $T_\mathrm{bath}$ with $I_\mathrm{clr}$~=~0. Similar behaviour can be observed also at other bath temperatures. The reason for the discrepancy remains unknown. Even though we can not assign any temperature for these thermometer curves, the shape of the \textit{I-V}s at e.g. $T_\mathrm{bath}$ = 1.5~K and $I_\mathrm{clr}$ = 10, 15 and 17~µA shows clear enhancement of the threshold current, indicating cooling.

\clearpage
\section{Cooler and thermometer \textit{I-V}s at $T_\mathrm{bath}$~=~1.4 to 2.4~K}
\label{supplementarysection:fulldata}
\begin{figure}[h!]
    \centering    \includegraphics[width = 11cm]{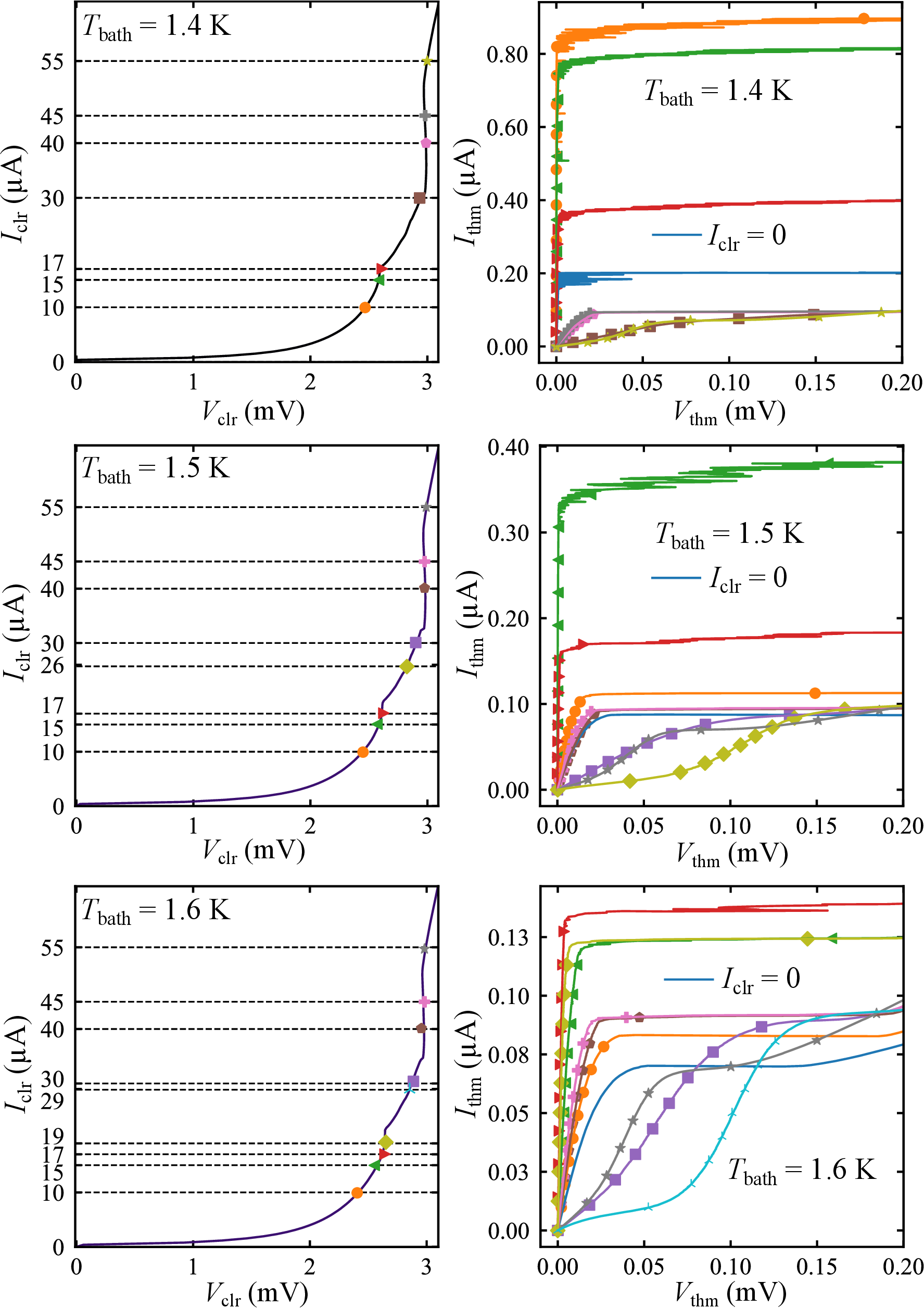}
    \caption{Cooler \textit{I-V}s with marked bias points and the corresponding thermometer \textit{I-V}s at $T_\mathrm{bath}$~=~1.4, 1.5 and 1.6~K.}
    \label{fig:supplementary_fulldata1.4to1.6}
\end{figure}

\begin{figure}[h!]
    \centering    \includegraphics[width = 11cm]{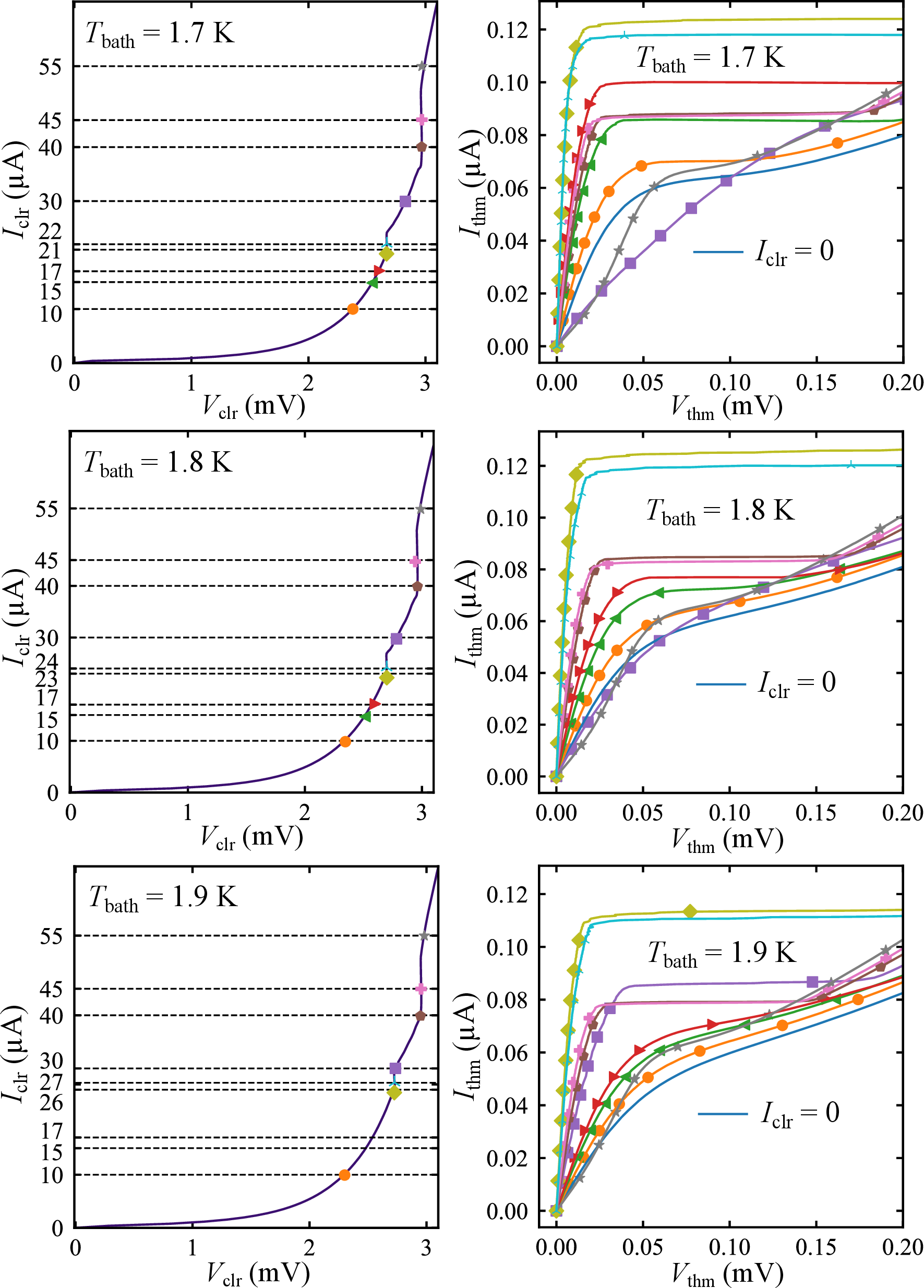}
    \caption{Cooler \textit{I-V}s with marked bias points and the corresponding thermometer \textit{I-V}s at $T_\mathrm{bath}$~=~1.7, 1.8 and 1.9~K.}
    \label{fig:supplementary_fulldata1.7to1.9}
\end{figure}

\begin{figure}[h!]
    \centering    \includegraphics[width = 11cm]{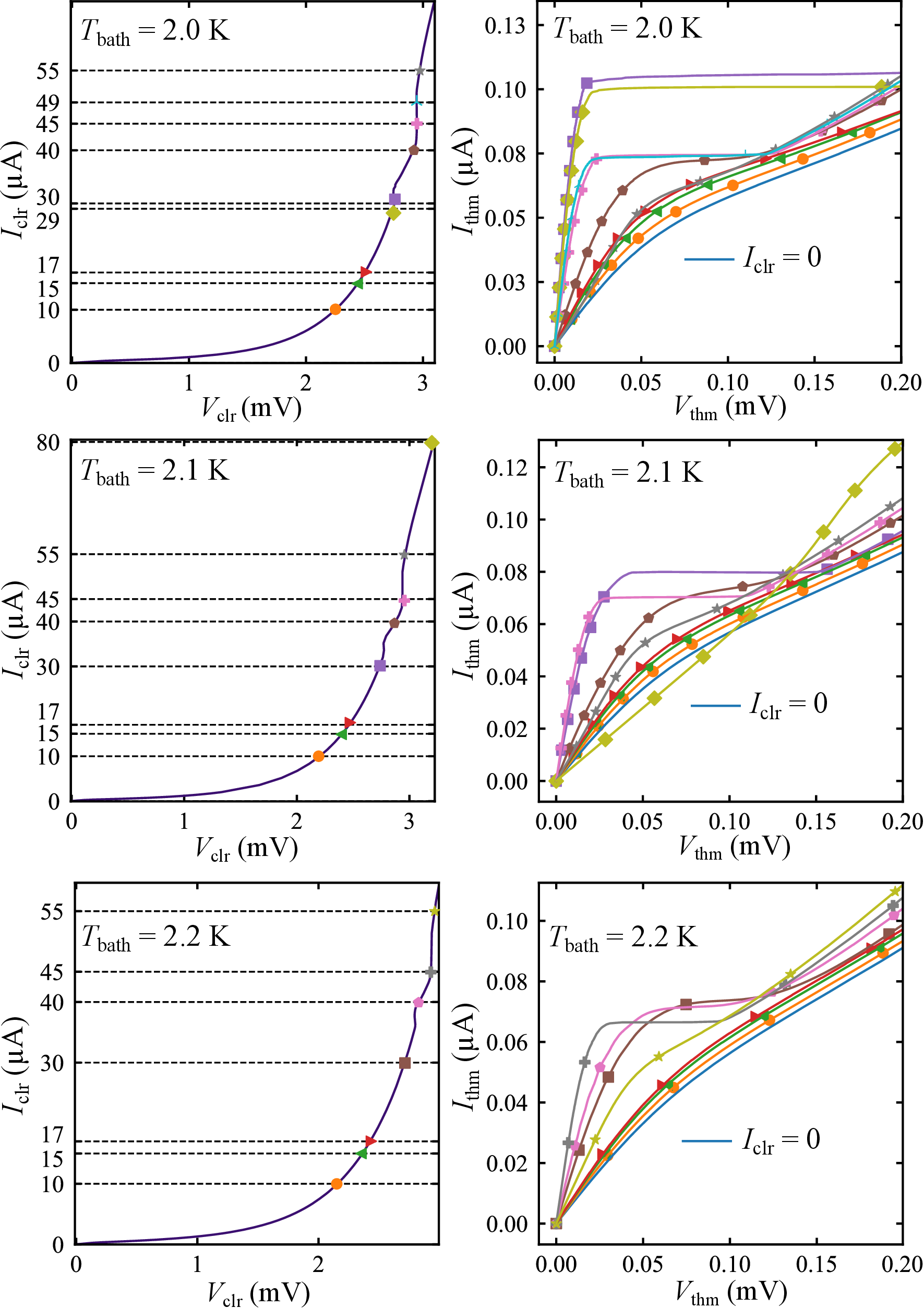}
    \caption{Cooler \textit{I-V}s with marked bias points and the corresponding thermometer \textit{I-V}s at $T_\mathrm{bath}$~=~2.0, 2.1 and 2.2~K.}
    \label{fig:supplementary_fulldata2.0K-2.2K}
\end{figure}

\begin{figure}[h!]
    \centering    \includegraphics[width = 11cm]{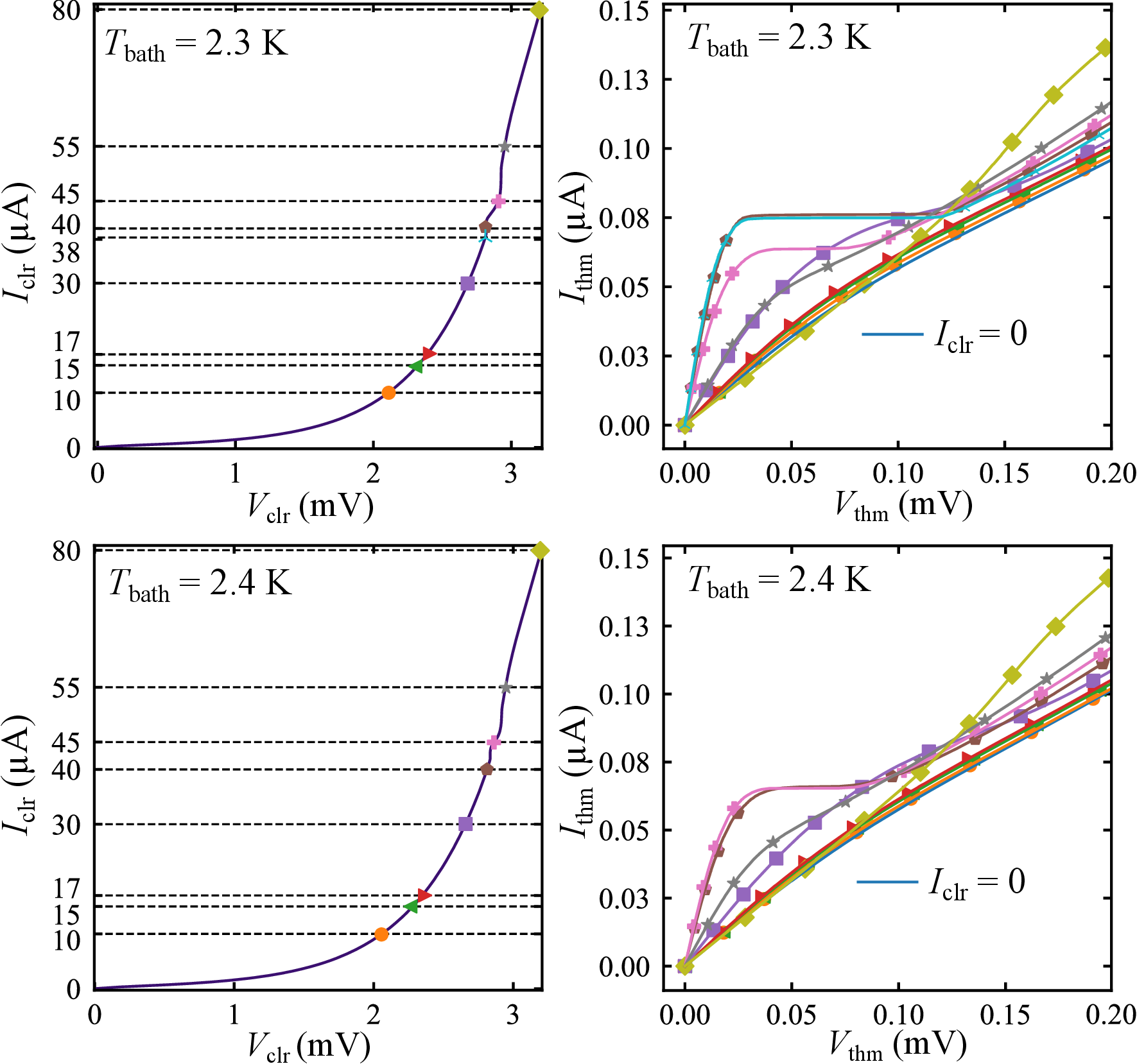}
    \caption{Cooler \textit{I-V}s with marked bias points and the corresponding thermometer \textit{I-V}s at $T_\mathrm{bath}$~=~2.3 and 2.4~K.}
    \label{fig:supplementary_fulldata2.3K-2.4K}
\end{figure}

Figures \ref{fig:supplementary_fulldata1.4to1.6}, \ref{fig:supplementary_fulldata1.7to1.9}, \ref{fig:supplementary_fulldata2.0K-2.2K} and \ref{fig:supplementary_fulldata2.3K-2.4K} show the cooling measurement data at each bath temperature between 1.4 and 2.4~K. At $T_\mathrm{bath}$~=~1.4, 1.5 and 1.6~K a supercurrent branch is observed at optimal cooling bias points.

\clearpage

\clearpage
\section{Finite element method simulations}
\label{supplementarysection:FEM}

\begin{figure}[h!]
    \centering
    \includegraphics[width = \textwidth]{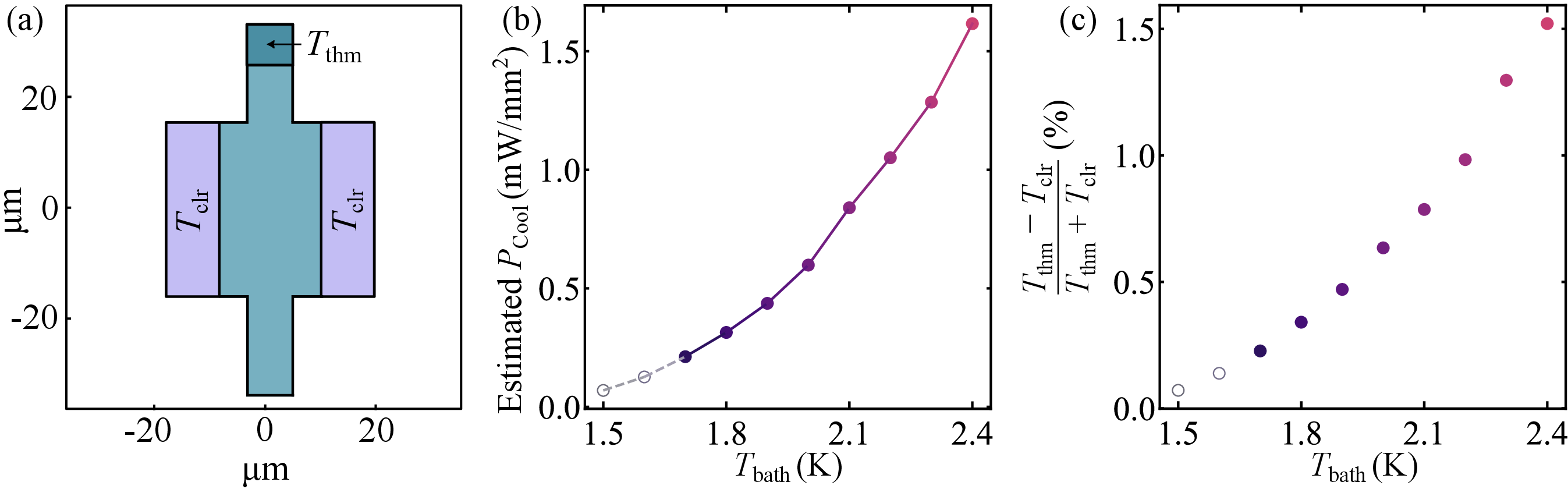}
    \caption{Finite element method simulation of the refrigerator. (a) Simulation geometry showing which areas the temperatures are averaged over. (b) Estimated cooling power density at optimal cooler bias for each bath temperature. (c) Relative difference between the thermometer and the cooler junction temperature as a function of the bath temperature. The model assumes normal state Al instead of superconducting state observed at $T_\mathrm{bath}$ up to 1.6~K at optimal cooler bias, thus the estimates at $T_\mathrm{bath}$~=~1.5 and 1.6~K is not expected to be reliable (see text S5).}
    \label{fig:supplementary_FEMl}
\end{figure}

The heat transport in normal state Al was modelled with a finite element method to obtain an estimate for the cooling power density and the temperature under the cooler junctions. The model gives estimates at higher bath temperatures but is expected to fail when the temperature of Al is at or below the critical temperature. The electron-electron thermal conductance and the electron-phonon coupling in Al are suppressed below the critical temperature due to the emergence of energy gap in the electron density of states, whereas our model assumes normal state Al. We used a two-dimensional model depicted in Fig.~\ref{fig:supplementary_FEMl}(a), since the thickness of Al is small compared to the other dimensions and relevant heat relaxation distances. A constant current density and cooling power were applied to the cooler junctions. Due to the finite electrical resistance of normal state Al (see Fig.~\ref{fig:supplementary_Altransition}) the current density induces Joule heating in the system. We expect the dominant heat relaxation method to be electron-phonon coupling. In temperature range relevant for our samples, the electron-phonon coupling constants for Al was obtained from Ref.~\cite{Santhanam1984InelasticFilms}. We expect phonons to be well thermalized to the bath. The temperatures at the thermometer and the cooler junctions were obtained as averages over the relevant areas.

Figure~\ref{fig:supplementary_FEMl}(b) shows the cooling power densities that would be consistent with the minimum temperature under the thermometer as extracted on the optimal cooler bias condition from the experimental dataset. Corresponding relative temperature differences at each $T_\mathrm{bath}$ are shown in Fig.~\ref{fig:supplementary_FEMl}(c).  

